\documentclass[twocolumn,prb]{revtex4}
\usepackage{graphicx}
\usepackage{dcolumn}
\usepackage{bm}
\usepackage{amsmath}
\usepackage{times}

\def\i{\mathrm{i}} 
\newcommand{\e}[1]{\mathrm{e}^{#1}}
\begin{document}

\preprint{}
\title{Heat transport by Dirac fermions in normal/superconducting graphene junctions}

\author{Takehito Yokoyama$^1$,  Jacob Linder$^2$ and Asle Sudb{\o}$^2$}
\affiliation{$^1$ Department of Applied Physics, Nagoya University, Nagoya, 464-8603, Japan%
\\
%and CREST, Japan Science and Technology Corporation (JST) Nagoya, 464-8603,
%Japan\\
$^2$ Department of Physics, Norwegian University of
Science and Technology, N-7491 Trondheim, Norway}

\date{\today}

\begin{abstract}
We study heat transport in normal/superconducting graphene junctions. 
We find that while the thermal conductance displays the usual exponential dependence on temperature, reflecting the $s$-wave symmetry of the superconductor, it exhibits an unusual oscillatory dependence on the potential height or the length of the barrier region. This oscillatory dependence stems from the emergent low-energy relativistic nature of fermions in graphene, essentially different from the result in conventional normal metal/superconductor junctions.
\end{abstract}

\maketitle

%--- title ---

%--- author ---

%
%--- address ---

%
%--- date ---

% It is always \today, today,
%  but any date may be explicitly specified
%-----------------------------------------------------------
%   Abstract
%-----------------------------------------------------------

%-----------------------------------------------------------

% PACS, the Physics and Astronomy
% Classification Scheme.
%\keywords{Suggested keywords}%Use showkeys class option if keyword
%display desired
%\section{Introduction}
The recent progress in practical fabrication techniques for a monoatomic layer of graphite, called graphene, has allowed for experimental studies of this system, which in turn has triggered a
tremendous interest \cite{Ando,Katsnelson,Castro,beenakker,novoselov,zhang,novoselov_nature}.
Graphene is a two-dimensional system of carbon atoms, and the low-energy electrons in graphene are governed by Dirac equation.
%, which connects condensed matter physics and quantum electrodynamics. 
Up to now, intensive studies on graphene have been conducted for instance quantum Hall effect\cite{zhang,Novoselov2,Yang}, minimum conductivity\cite{novoselov_nature} and bipolar supercurrent\cite{heersche}.
\par
From applied physics point of view, graphene is also an important material. Graphene exhibits high mobility and carrier density controllable by gate voltage,  which makes it well suited for achieving device applications.\cite{novoselov,zhang,Bunch,Berger} In order to apply graphene to electric devices, it is an important issue to clarify characteristics of transport phenomena in graphene. 
\par
In conventional normal metal/superconductor junctions, it is known that electric and thermal conductances reflect the magnitude or symmetry of the gap of the superconductor.\cite{btk,Andreev} While conductance in normal/superconductor graphene junction has been studied,\cite{beenakker2,sengupta,linder} thermal conductance in the same junction has not yet been investigated. The study of the thermal conductance in normal/superconductor graphene junction will complement the study of the conductance in the same junction. 

In this paper, we study heat transport in normal/superconducting graphene junctions. 
We find that the thermal conductance has an exponential dependence on temperature, which reflects the $s$-wave symmetry of the superconductor. However, it displays an oscillatory dependence on the potential height or the length of the barrier region. This oscillatory dependence stems from the relativistic nature of fermions in graphene, and differs in an essential way from the result in the conventional normal metal/superconductor junctions. 

\begin{figure}[htb]
\begin{center}
\scalebox{0.4}{
\includegraphics[width=19.0cm,clip]{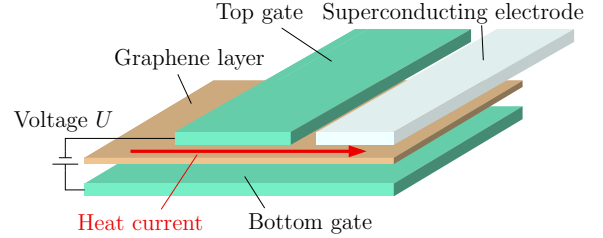}}
\end{center}
\caption{(color online) The proposed experimental setup to measure heat transport by Dirac fermions in a graphene normal/superconductor proximity structure. The top and bottom gate allow for the chemical potential in the middle region to be adjusted.} \label{fig:model}
\end{figure}

%\section{Formulation}
We briefly present the formalism to be used in this paper, following Ref. \cite{linder}. 
Consider a two dimensional normal/insulating/superconducting graphene junction\cite{note} where the superconducting (normal) region is located in the semi-infinite regions $x>L$ $(x<0)$. The proposed experimental setup of our model is shown in Fig. \ref{fig:model}. 
By exploiting the valley degeneracy,\cite{morpurgo2} the Bogoliubov-de Gennes equation for the junction in the $xy$-plane reads 
\begin{eqnarray}\label{eq:Bdg2}
\begin{pmatrix}
 H - E_{F} \hat{1} & \Delta \hat{1} \\
\Delta^\dag \hat{1} & E_{F}\hat{1} - H \\
\end{pmatrix}
\begin{pmatrix}
u\\
v\\
\end{pmatrix}
 = E \begin{pmatrix}
 u\\
 v\\
 \end{pmatrix}
\end{eqnarray}
with $H = v_F (k_x \sigma_x + k_y \sigma_y) $. The superconducting order parameter reads 
\begin{equation}
\Delta=\Delta(T) \e{i\phi} \Theta(x-L),
\end{equation}
where $\Theta(x)$ is the Heaviside step function, while $\phi$ is the phase corresponding to the globally broken $U(1)$ symmetry in the superconductor. 
Also, $v_F \approx 10^{6}$m/s  is the energy-independent Fermi velocity for graphene, $\sigma_i (i=x,y)$ denotes the Pauli matrices, $E$ is the excitation energy, and $u$ and $v$ denote the electron-like and hole-like exictations, respectively, described by the wave-function. 
The Pauli matrices operate on the two triangular sublattice space of the
honeycomb structure, corresponding to the A and B atoms. 
The linear dispersion relation is a reasonable approximation even for Fermi levels
as high as 1 eV,\cite{wallace} such that the fermions in graphene behave
like massless Dirac fermions in the low-energy regime.
\par
Let us consider an incident electron from the normal side of the junction $(x<0)$ with energy $E$. For positive excitation energies $E>0$, the eigenvectors and corresponding momentum of the particles read 
\begin{eqnarray}
\psi^{e}_+ &= [1, \e{\i\theta}, 0,0]^\mathcal{T}\e{\i p^{e}\cos\theta x},\; p^{e} = (E +  E_{F})/v_{F},
\end{eqnarray}
for a right-moving electron at angle of incidence $\theta$, while a left-moving 
electron is described by the substitution $\theta\to\pi-\theta$. The superscript $^\mathcal{T}$ denotes the transpose. If Andreev-reflection takes place, a left-moving 
hole with energy $E$ and angle of reflection $\theta_{A}$ is generated with corresponding wave-function
\begin{eqnarray}
\psi^{h}_- = [0,0,1,\e{-\i\theta_{A}}]^\mathcal{T}\e{-\i p^{h}\cos\theta_{A} x},\; p^{h} = (E - E_{F})/v_{F},
\end{eqnarray}
where the superscript e (h) denotes an electron-like (hole-like) excitation. Since translational invariance in 
the $y$-direction holds, the corresponding component of momentum is conserved. This condition 
allows for determination of the Andreev-reflection angle $\theta_{A}$ via $p^{h}\sin\theta_{A} = p^{e}\sin\theta.$ From this equation, one infers that there is no Andreev-reflection ($\theta_{A} = \pm\pi/2$) for angles of incidence above the critical angle 
\begin{equation}
\theta_{c} = \sin^{-1} ( |E-E_{F}|/(E+E_{F}) ).
\end{equation}
On the superconducting side of the system ($x>L$), the possible wavefunctions for transmission of a right-moving quasiparticle with a given excitation energy $E>0$ reads
\begin{eqnarray}\label{eq:psisuper}
\Psi^{e}_+ = \Big(u, u\e{\i\theta^+},v\e{-\i\phi},v\e{\i(\theta^+-\phi)}\Big)^{T}\nonumber\\
\times\e{\i q^{e}\cos\theta^+ x},\; q^{e} = (E'_{F} + \sqrt{E^2-\Delta^2})/v_{F},\\
\Psi^{h}_- = \Big(v, v\e{\i\theta^-},u\e{-\i\phi},u\e{\i(\theta^--\phi)}\Big)^{T}\nonumber\\
\times\e{\i q^{h}\cos\theta^- x},\; q^{h} = (E'_{F} - \sqrt{E^2-\Delta^2})/v_{F}.
\end{eqnarray}
The coherence factors are given by \cite{sudbo}
\begin{eqnarray}
u = \sqrt{\frac{1}{2}\Big(1 + \frac{\sqrt{E^2-|\Delta|^2}}{E}\Big)}, \\
v = \sqrt{\frac{1}{2}\Big(1 - \frac{\sqrt{E^2-|\Delta|^2}}{E}\Big)}.
\end{eqnarray}
Above, we have defined $\theta^+ = \theta_{S}^{e}$ and $\theta^- = \pi-\theta_{S}^{h}$. 
The transmission angles $\theta^{(i)}_{S}$ for the electron-like and hole-like quasiparticles 
are given by $q^{(i)}\sin\theta^{(i)}_{S} = p^{e} \sin\theta$, i$=$e,h. Note that in all the wavefunctions listed above, for clarity we have not included a common phase factor $\e{\i k_y y}$ which corresponds to the conserved momentum in the $y$-direction. 
\par
It is appropriate to insert the restriction which will be used throughout the paper, namely $\Delta \ll E_{F}'$. Since we are using a mean-field approach to describe the superconducting part of the Hamiltonian, phase-fluctuations of the order parameter have to be small \cite{phase-fluctuations}.
\par
We define the wavefunctions in the normal, insulating and superconducting regions by $\psi$, $\tilde{\psi}_{I}$ and $\Psi$, respectively, with
\begin{eqnarray}
\psi = \psi^{e}_+ + r\psi^{e}_- + r_{A}\psi^{h}_-, \\
\tilde{\psi}_{I} = \tilde{t}_1\tilde{\psi}^{e}_+ + \tilde{t}_2\tilde{\psi}^{e}_- + \tilde{t}_3\tilde{\psi}^{h}_+ + \tilde{t}_4\tilde{\psi}^{h}_-, \\
\Psi =  t^{e}\Psi^{e}_+ + t^{h}\Psi^{h}_-.
\end{eqnarray}
The wavefunctions $\tilde{\psi}$ differ from $\psi$ in that the Fermi energy is  shifted by an external potential, such that $E_{F} \to E_{F}-U$ where $U$ is the barrier height. Also, note that the trajectories of the quasiparticles in the insulating region, defined by the angles $\tilde{\theta}$ and $\tilde{\theta}_{A}$, differ by the same substitution:
\begin{eqnarray}
\sin\tilde{\theta}/\sin\theta = (E+E_{F})/(E+E_{F}-U),\\
\sin\tilde{\theta}_{A}/\sin\theta = (E+E_{F})/(E-E_{F}+U).
\end{eqnarray}
Note that the subscript $\pm$ on the wavefunctions indicates the direction of momentum, which is in general different from the group velocity direction. 
\par
By matching the wavefunctions at both interfaces, 
%\begin{eqnarray}\label{eq:nisboundary}
$\psi|_{x=0} = \tilde{\psi}_ {I}|_{x=0}$ and $\tilde{\psi}_{I}|_{x=L} = \Psi |_{x=L}$ ,\cite{bc}
%\end{eqnarray}
we obtain the following expressions for the normal reflection coefficient $r$ and the Andreev-reflection coefficient $r_A$: \cite{linder}
\begin{eqnarray}
r = t_e(A+C) + t_h(B+D) - 1, \\
r_A = t_e(A'+C') + t_h(B'+D'),
\end{eqnarray}
where the transmission coefficients read
\begin{widetext}
\begin{align}
t_e &= 2\cos\theta[\e{-\i\theta_A}(B'+D') - (B'\e{-\i\tilde{\theta}_A} - D'\e{\i\tilde{\theta}_A})]\rho^{-1}, \\
t_h &= t_e[\e{\i\theta_A}(A'\e{-\i\tilde{\theta}_A} - C'\e{\i\tilde{\theta}_A})-A'-C'] [B'+D'-\e{\i\theta_A}(B'\e{-\i\tilde{\theta}_A}-D'\e{\i\tilde{\theta}_A})]^{-1}, \\
\rho &= [ \e{-\i\theta_A}(B'+D') - (B'\e{-\i\tilde{\theta}_A} -
D'\e{\i\tilde{\theta}_A})][ \e{-\i\theta}(A+C) +
(A\e{\i\tilde{\theta}} - C\e{-\i\tilde{\theta}})] \nonumber\\
&-[(D\e{-\i\tilde{\theta}} - B\e{\i\tilde{\theta}}) -
\e{-\i\theta}(B+D)][A'\e{-\i\tilde{\theta}_A}-C'\e{\i\tilde{\theta}_A} -
\e{-\i\theta_A}(A'+C')]
\end{align}
\end{widetext}
and we have introduced the auxiliary quantities
\begin{eqnarray}
A = u \e{\i(q^+-p^+)}[1 - (\e{\i\tilde{\theta}} - \e{\i\theta^+})(2\cos\tilde{\theta})^{-1}],\nonumber\\
B = v \e{\i(q^--p^+)}[1 - (\e{\i\tilde{\theta}} - \e{\i\theta^-})(2\cos\tilde{\theta})^{-1}],\nonumber\\
C = u \e{\i(p^++q^+)}(\e{\i\tilde{\theta}} - \e{\i\theta^+})(2\cos\tilde{\theta})^{-1},\nonumber\\
D = v \e{\i(p^++q^-)}(\e{\i\tilde{\theta}} - \e{\i\theta^-})(2\cos\tilde{\theta})^{-1},
\end{eqnarray}
\begin{eqnarray}
A' = v \e{\i(q^++p^- - \phi)}[1 + (\e{\i\theta^+} - \e{-\i\tilde{\theta}_A})(2\cos\tilde{\theta}_A)^{-1}],\nonumber\\
B' = u \e{\i(q^-+p^- - \phi)}[1 + (\e{\i\theta^-} - \e{-\i\tilde{\theta}_A})(2\cos\tilde{\theta}_A)^{-1}],\nonumber\\
C' = v \e{\i(q^+-p^--\phi)}(\e{-\i\tilde{\theta}_A} - \e{\i\theta^+})(2\cos\tilde{\theta}_A)^{-1},\nonumber\\
D' = u \e{\i(q^--p^--\phi)}(\e{-\i\tilde{\theta}_A} - \e{\i\theta^-})(2\cos\tilde{\theta}_A)^{-1}.
\end{eqnarray}
Here, we have defined
\begin{align}
q^+ &= q^e \cos \theta^+ L, \;\; q^- = q^h \cos\theta^- L, \notag\\
p^+ &= \tilde{p}^e \cos\tilde{\theta} L, \;\; p^- = \tilde{p}^h \cos\tilde{\theta}_A L.
\end{align}
In the thin-barrier limit defined as $L \to 0$ and $U \to\infty$, one gets 
\begin{eqnarray}
\tilde{\theta} \to 0,\; \tilde{\theta}_A \to 0,\; q_\pm\to 0, p_\pm \to \mp\chi
\end{eqnarray}
with $\chi = LU/v_F$. This indicates that thermal conductance is $\pi$-periodic with respect to $\chi$ in this limit. 
\par
Finally, the normalized thermal conductance is given by 
\begin{widetext}
\begin{eqnarray}
\kappa  = \int_{ 0 }^\infty  {\int_{ - \pi /2}^{\pi /2} {dEd\theta \cos \theta (1 - \left| {r(E,\theta )} \right|^2  - {\mathop{\rm Re}\nolimits} (\frac{{\cos \theta _A }}{{\cos \theta }})\left| {r_A (E,\theta )} \right|^2 )\frac{{E^2 }}{{\Delta _0 T^2 \cosh ^2 (\frac{E}{{2T}})}}} } 
\end{eqnarray}
\end{widetext}
with the gap at zero temperature $\Delta _0 \equiv \Delta(0) $.

%%%%%%%%%%%%%%%%%%%%%%%%%%%%%%%%
%\section{Results}
\begin{figure}[htb]
\begin{center}
\scalebox{0.4}{
\includegraphics[width=19.5cm,clip]{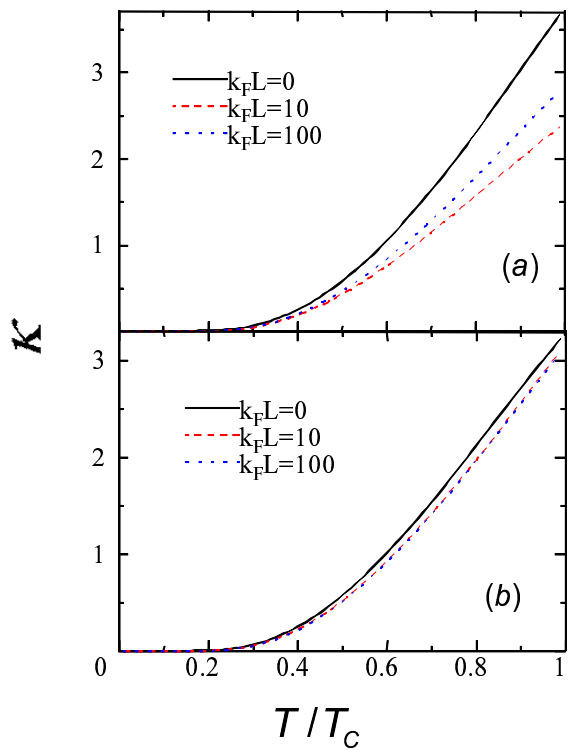}}
\end{center}
\caption{(Color online) Thermal conductance as a function of $T/T_C$ for various $k_F L$ with $U/E_F=10$ and $E'_F=100 \Delta_0$ at $E_F=100 \Delta_0$ in (a) and $E_F=10 \Delta_0$ in (b). } \label{f1}
\end{figure}

\begin{figure}[htb]
\begin{center}
\scalebox{0.4}{
\includegraphics[width=19.5cm,clip]{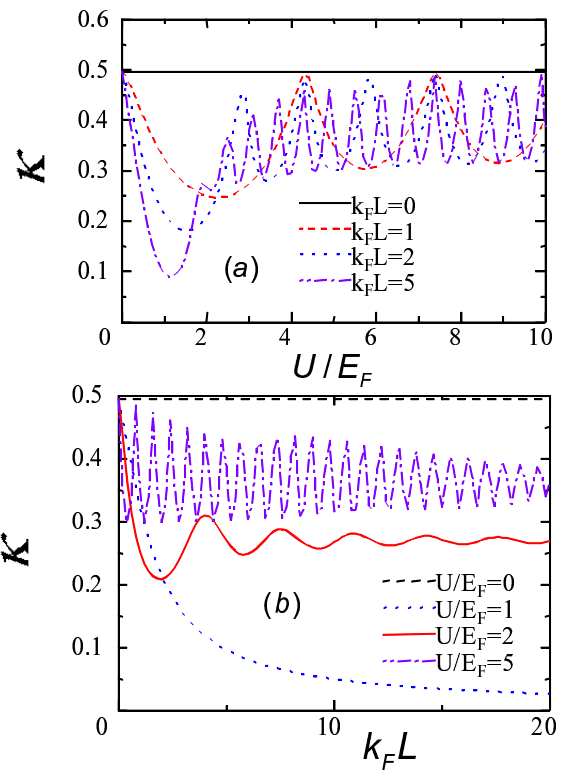}}
\end{center}
\caption{(Color online) (a) Thermal conductance as a function of $U/E_F$ for various $k_F L$ with $T/T_C=0.5$ and $E_F=E'_F=100 \Delta_0$.
 (b)  Thermal conductance as a function of $k_F L$ for various $U/E_F$ with $T/T_C=0.5$ and $E_F=  E'_F=100 \Delta_0$. } \label{f2}
\end{figure}

We next present our results for the normalized thermal conductance. 
 Figure \ref{f1} (a) shows thermal conductance as a function of $T/T_C$ for various $k_F L$ with $U/E_F=10$ and $E_F=E'_F=100 \Delta_0$. Here $T_C$ is the transition temperature and  $k_F \equiv E_F/v_F$. From Fig. \ref{f1} 
 (a), an exponential dependence of the thermal conductance on temperature is seen, similar to the conventional 
 normal metal/superconductor junctions.\cite{Andreev} This exponential dependence reflects the $s$-wave symmetry 
 of the superconductor. However, the length dependence of the thermal conductance is nonmonotonic (oscillatory) 
 and thus essentially different from that in the conventional normal metal/superconductor junctions. 
 A similar plot for $E_F=10 \Delta_0$ is shown in Fig. \ref{f1} (b). We also find an exponential temperature dependence, but the dependence on $L$ gets weaker. 
Therefore, the magnitude of the oscillation with respect to $k_F L$ gets reduced with the increase of the Fermi wave vector mismatch. 

 Figure \ref{f2} (a) depicts thermal conductance as a function of $U/E_F$ for various length $k_F L$ with $T/T_C=0.5$ and $E_F=E'_F=100 \Delta_0$. An oscillatory dependence of the thermal conductance on $U/E_F$ is seen. The period decreases with $k_F L$. 
 Figure \ref{f2} (b) displays thermal conductances as a function of $k_F L$ for various $U/E_F$ with $T/T_C=0.5$ and $E_F=  E'_F=100 \Delta_0$. We also find an oscillatory dependence on $k_F L$. The period also decreases with $U/E_F$. 
These features stem from the $\pi$-periodicity of the thermal conductance with respect to $\chi=k_F L U/E_F$ in the limit of $U \gg E_F$ and  $k_F L \ll 1$ similar to the junction conductance.\cite{sengupta,linder,katsnelson}
In other words, the damped oscillatory behavior of the thermal conductance is a direct manifestation of the relativistic low-energy Dirac fermions. Also, the presence of the insulating region is essential for the oscillatory behavior. 
%Thus, the reduced magnitude of the oscillation with the increase of the Fermi wave vector mismatch as seen from Fig. \ref{f1} is understandable because the increase of the Fermi wave vector mismatch enhances normal reflection probability and hence the presence of the insulating region becomes less vital. 
\par
Since we have assumed a homogeneous chemical potential in each of graphene regions, the experimental observation of the 
predicted effects require charge homogeneity of the graphene samples. This is a challenge, since electron-hole puddles in graphene 
imaged by a scanning single electron transistor device \cite{puddles} suggest that such charge inhomogeneities play an important 
role in limiting the transport characteristics of graphene \cite{kim}. In addition, we have neglected the spatial variation of the 
superconducting gap near the interface. The suppression of the order parameter near the interface is expected to be least pronounced when the sharp 
edge criteria is satisfied and there is a large Fermi-vector mismatch. In the present case, this is 
precisely so, whence we do not expect our qualitative results to be affected by taking into account the reduction of the gap near the 
interface. Finally, we have assumed that there is no lattice mismatch at the interfaces and that these are smooth and impurity-free \cite{beenakker}. A more refined picture could be obtained by using more realistic models of the variation of the chemical potential, i.e. a continuous slope instead of a step-like variation. 
\par
In summary, we have studied heat transport in normal/superconducting graphene junctions. 
We found that the thermal conductance has an exponential dependence on temperature which reflects the $s$-wave symmetry of the superconductor but oscillatory dependence on the potential height or the length of the barrier region. This oscillatory dependence stems from the relativistic nature of fermions in graphene, essentially different from the result in the conventional normal metal/superconductor junctions.

%==================================================
T.Y. acknowledges support by the JSPS. 
J.L. and A.S. were supported by the Research Council of Norway, Grants 
No. 158518/431 and No. 158547/431 (NANOMAT), and Grant
No. 167498/V30 (STORFORSK).
%This work was supported by
%NAREGI Nanoscience Project, the Ministry of Education, Culture,
%Sports, Science and Technology, Japan, the Core Research for Evolutional
%Science and Technology (CREST) of the Japan Science and Technology
%Corporation (JST) and a Grant-in-Aid for the 21st Century COE "Frontiers of
%Computational Science" . The computational aspect of this work has been
%performed at the Research Center for Computational Science, Okazaki National
%Research Institutes and the facilities of the Supercomputer Center,
%Institute for Solid State Physics, University of Tokyo and the Computer Center.
%======Reference===================================
%

%===============================================================


\begin{thebibliography}{99}
\bibitem{Ando} T. Ando, J. Phys. Soc. Jpn. \textbf{74}, 777 (2005).

\bibitem{Katsnelson} M. I. Katsnelson and K. S. Novoselov, Solid State Commun. \textbf{143}, 3 (2007).

\bibitem{Castro} A. H. Castro Neto, F. Guinea, N. M. R. Peres, K. S. Novoselov and A. K. Geim, arXiv:0709.1163v1.

\bibitem{beenakker} C. W. J. Beenakker, arXiv:0710.3848.

%\bibitem{Fradkin} E. Fradkin, \textit{Field Theories of Condensed Matter Systems} (Westview, Oxford,1997).

%\bibitem{Volovik} G. E. Volovik, \textit{ The Universe in a Helium Droplet} (Clarendon, Oxford, 2003).

\bibitem{novoselov}  K. S. Novoselov, A. K. Geim, S. V. Morozov, D. Jiang, Y. Zhang, S. V. Dubonos, I. V. Grigorieva and A. A. Firsov, Science \textbf{306}, 666 (2004).

\bibitem{zhang} Y. Zhang, Y.-W. Tan, H. L. Stormer and P. Kim , Nature \textbf{438}, 201 (2005).

\bibitem{novoselov_nature} K. S. Novoselov, A. K. Geim, S. V. Morozov, D. Jiang, M. I. Katsnelson, I. V. Grigorieva, S. V. Dubonos, and A. A. Firsov, Nature \textbf{438}, 197 (2005).

\bibitem{Novoselov2} K. S. Novoselov, E. McCann, S. V. Morozov, V. I. Fal'ko, M. I. Kastenelson, U. Zeitler, D. Jiang, F. Schedin, and A. K. Geim, Nat. Phys. \textbf{2}, 177 (2006).

\bibitem{Yang} K. Yang, Solid State Commun. \textbf{143}, 27 (2007).

\bibitem{heersche}  H. B. Heersche, P. Jarillo-Herrero, J. B. Oostinga, L. M. K. Vandersypen and A. F. Morpurgo, Nature \textbf{446}, 56 (2007).

\bibitem{Bunch}
J. Scott Bunch, Y. Yaish, M. Brink, K. Bolotin, and P. L. McEuen, Nano Lett. \textbf{5}, 287 (2005).

\bibitem{Berger}
C. Berger, Z. Song, X. Li, X. Wu, N. Brown, C. Naud, D. Mayou, T. Li, J. Hass, A. N. Marchenkov, E. H. Conrad, P. N. First, and W. A. de Heer, Science \textbf{312}, 1191 (2006).

%\bibitem{kasumov} A. Yu. Kasumov, R. Deblock, M. Kociak, B. Reulet, H. Bouchiat, I. I. Khodos, Yu. B. Gorbatov, V. T. Volkov, C. Journet, M. Burghard, Science \textbf{284}, 1508 (1999).

%\bibitem{morpurgo} A. F. Morpurgo, J. Kong, C. M. Marcus, H. Dai, Science \textbf{286}, 263 (1999).

%\bibitem{Schmidt} G. Schmidt, D. Ferrand, L. W. Molenkamp, A. T. Filip and B. J. van Wees, Phys. Rev. B \textbf{62}, R4790 (2000).

%\bibitem{Kane} C. L. Kane and E. J. Mele, Phys. Rev. Lett. \textbf{95}, 226801 (2005). 

%\bibitem{Hernando} D. Huertas-Hernando, F. Guinea, and A. Brataas, Phys. Rev. B \textbf{74}, 155426 (2006).

%\bibitem{Tombros} N. Tombros, C. Jozsa, M. Popinciuc, H. T. Jonkman, and B. J. van Wees, Nature (London) \textbf{448}, 571 (2007). 

\bibitem{Andreev} A. F. Andreev, Sov. Phys. JETP \textbf{19}, 1228 (1964). 

\bibitem{btk} G. E. Blonder, M. Tinkham, and T. M. Klapwijk, 
Phys. Rev. B {\bf 25}, 4515 (1982).

\bibitem{beenakker2} C. W. J. Beenakker, Phys. Rev. Lett. \textbf{97}, 067007 (2006).

\bibitem{sengupta} S. Bhattacharjee and K. Sengupta, Phys. Rev. Lett. \textbf{97}, 217001 (2006); S. Bhattacharjee, M. Maiti, and K. Sengupta, Phys. Rev. B \textbf{76}, 184514 (2007).

\bibitem{linder} J. Linder and A. Sudb{\o}, Phys. Rev. Lett. \textbf{99}, 147001 (2007); arXiv:0712.0831.

\bibitem{note} We underline that the notation "insulator" in this context refers to a normal segment of graphene in which one experimentally induces an effective potential barrier.

\bibitem{morpurgo2} A. F. Morpurgo and F. Guinea, Phys. Rev. Lett. \textbf{97}, 196804 (2006).

\bibitem{wallace} P. R. Wallace, Phys. Rev. \textbf{71}, 622 (1947).

\bibitem{sudbo} K. Fossheim and A. Sudb{\o}, {\it Superconductivity: Physics and applications}, John Wiley \& Sons Ltd., Ch. 5 (2004).

\bibitem{phase-fluctuations} A. K. Nguyen and A. Sudb{\o}, Phys. Rev. B {\bf 60}, 15307 (1999);
H. Kleinert, Phys. Rev. Lett., {\bf 84} 286 (2000).

\bibitem{bc}
Note that these conditions are equivalent to $\hat v_x \psi|_{x=0} = \hat v_x \tilde{\psi}_ {I}|_{x=0}$ and $\hat v_x \tilde{\psi}_{I}|_{x=L} = \hat v_x \Psi |_{x=L}$ with velocity operator $\hat v_x  = \partial H /\partial k_x  = v_F\sigma _x$ and hence the current is conserved at the interfaces. 

\bibitem{katsnelson} M. I. Katsnelson, K. S. Novoselov and A. K. Geim, Nature Phys. \textbf{2}, 620 (2006).

\bibitem{puddles}  J. Martin, N. Akerman, G. Ulbricht, T. Lohmann, J. H.
Smet, K. von Klitzing and A. Yacoby, arXiv:0705.2180v1.
% Nat. Phys. {\bf 3}, XXX (2007);

\bibitem{kim} E.-A. Kim and A. H. Castro Neto, arXiv:cond-mat/0702562v2.

\end{thebibliography}
\end{document}